\documentstyle[onecolumn]{mn}

\def\beq{\begin{equation}}
\def\eeq{\end{equation}}
\def\bea{\begin{eqnarray}}
\def\eea{\end{eqnarray}}
\def\beas{\begin{eqnarray*}}
\def\eeas{\end{eqnarray*}}
\def\non{\nonumber}
\def\va{{\vec a}}
\def\vd{{\vec d}}
\def\vg{{\vec g}}
\def\vh{{\vec h}}
\def\vk{{\vec k}}
\def\vm{{\vec m}}
\def\vn{{\vec n}}
\def\vq{{\vec q}}
\def\vs{{\vec s}}
\def\vsigma{{\vec\sigma}}
\def\vt{{\vec t}}
\def\vu{{\vec u}}
\def\vG{{\vec G}}
\def\vJ{{\vec J}}
\def\vP{{\vec P}}
\def\vT{{\vec T}}
\def\tg{{\tilde g}}
\def\ttg{{\tilde\tg}}
\def\vgt{{\tilde\vg}}
\def\vgtt{{\tilde\vgt}}

\def\cD{{\cal D}}

\def\cF{{\cal F}}
\def\cG{{\cal G}}
\def\cH{{\cal H}}

\def\cK{{\cal K}}

\def\cM{{\cal M}}
\def\cN{{\cal N}}
\def\cO{{\cal O}}
\def\cR{{\cal R}}
\def\cY{{\cal Y}}
\def\cZ{{\cal Z}}

\def\tcR{{\tilde\cR}}
\def\tcRh{{\tcR_{\rm h}}}

\def\cU{{\cal U}}
\def\vph{{v^{(\hat\phi)}}}
\def\vrh{{v^{(\hat r)}}}
\def\vrb{{v^{(\bar r)}}}
\def\lam{\lambda}
\def\lamk{\lam_{\rm k}}
\def\lamms{\lam_{\rm ms}}
\def\lamph{\lam_{\rm ph}}
\def\lams{\lambda_{\rm s}}
\def\lamout{\lam_{\rm out}}
\def\Es{E_{\rm s}}
\def\Em{E_{\rm m}}
\def\Emin{{E_{\rm min}}}
\def\Emax{{E_{\rm max}}}
\def\dM{\dot{M}}
\def\dcM{\dot{\cM}}
\def\dcMm{\dcM_{\rm m}}
\def\mGamma{{\mit\Gamma}}
\def\mSigma{{\mit\Sigma}}
\def\Omegak{\Omega_{\rm k}}
\def\Omegas{\Omega_{\rm s}}
\def\tOmega{{\tilde\Omega}}
\def\gam{{\gamma}}
\def\gamk{{\gam_{\rm k}}}
\def\gamt{{\gam_{\rm t}}}
\def\gamh{{\gam_{\rm h}}}

\def\rin{r_{\rm in}}
\def\rout{r_{\rm out}}
\def\rh{r_{\rm h}}
\def\rg{r_{\rm g}}
\def\rms{r_{\rm ms}}
\def\rph{r_{\rm ph}}
\def\rs{r_{\rm s}}
\def\rsin{r_{\rm s,in}}
\def\rsint{r_{\rm s,int}}
\def\rsout{r_{\rm s,out}}
\def\cs{c_{\rm s}}
\def\cad{c_{\rm ad}}
\def\lh{l_{\rm h}}
\def\lk{l_{\rm k}}
\def\Lh{L_{\rm h}}
\def\uh{u_{\rm h}}
\def\nuh{\nu_{\rm h}}
\def\cZh{\cZ_{\rm h}}
\def\hot{type I }
\def\Hot{Type I }
\def\cold{type II }
\def\Cold{Type II }
\def\outer{{{\it outer} }}

\def\inter{{{\it intermediate} }}
\def\inner{{{\it inner} }}
\def\Inner{{{\it Inner} }}
\def\vec#1{\ifmmode\mathchoice{\mbox{\boldmath$\displaystyle#1$}}
{\mbox{\boldmath$\textstyle#1$}}
{\mbox{\boldmath$\scriptstyle#1$}}
{\mbox{\boldmath$\scriptscriptstyle#1$}}\else
\hbox{\boldmath$\textstyle#1$}\fi}
\def\ga{\mathrel{\mathchoice {\vcenter{\offinterlineskip\halign{\hfil
$\displaystyle##$\hfil\cr>\cr\sim\cr}}}
{\vcenter{\offinterlineskip\halign{\hfil$\textstyle##$\hfil\cr
>\cr\sim\cr}}}
{\vcenter{\offinterlineskip\halign{\hfil$\scriptstyle##$\hfil\cr
>\cr\sim\cr}}}
{\vcenter{\offinterlineskip\halign{\hfil$\scriptscriptstyle##$\hfil\cr
>\cr\sim\cr}}}}}
\def\la{\mathrel{\mathchoice {\vcenter{\offinterlineskip\halign{\hfil
$\displaystyle##$\hfil\cr<\cr\sim\cr}}}
{\vcenter{\offinterlineskip\halign{\hfil$\textstyle##$\hfil\cr
<\cr\sim\cr}}}
{\vcenter{\offinterlineskip\halign{\hfil$\scriptstyle##$\hfil\cr
<\cr\sim\cr}}}
{\vcenter{\offinterlineskip\halign{\hfil$\scriptscriptstyle##$\hfil\cr
<\cr\sim\cr}}}}}

\begin{document}

\title{Viscous accretion discs around rotating black holes}

\author[J. Peitz and S. Appl]
{Jochen Peitz$^1$\thanks{e-mail: {\tt jpeitz@lsw.uni-heidelberg.de}}
and Stefan Appl$^2$\thanks{e-mail: {\tt appl@astro.u-strasbg.fr}}\\
$^1$Landessternwarte K\"onigstuhl, 69117 Heidelberg, Germany\\
$^2$Observatoire Astronomique, Universit\'{e} Louis Pasteur,
11, rue de l'Universit\'{e}, 67000 Strasbourg, France}

\maketitle

\begin{abstract}
The stationary hydrodynamic equations for transonic viscous accretion discs in Kerr geometry are derived. 
The consistent formulation is given for the viscous angular momentum transport and the
boundary conditions on the horizon of a central black hole.
An expression for the thickness of the disc is obtained from the vertical Euler equation
for general accretion flows with vanishing vertical velocity.
Different solution topologies are identified, characterized by a sonic transition close to or far from
the marginally stable orbit.
A numerical method is presented that allows to integrate the structure equations of transonic 
accretion flows.
Global polytropic solutions for the disc structure are calculated, covering
each topology and a wide range of physical conditions.
These solutions generally possess a sub-Keplerian angular momentum distribution and have maximum temperatures
in the range $10^{11}-10^{12}$ K.
Accretion discs around rotating black holes are hotter and deposit less angular momentum on the 
central object than accretion discs around Schwarzschild black holes.
\end{abstract}


%
\section{Introduction}
\label{sec1}
The presence of hot optically thin gas in accreting black hole systems such as X-ray 
binaries or active galactic nuclei has been established by X-ray and $\gamma$\/-ray observations.
The spectra are very hard, with power-law photon indices $\alpha_{\rm N}\sim1.5-2.5$\/
in the energy range $E\sim$\/ keV to 100 keV
(Gilfanov et al. 1995; Maisack et al. 1993; Johnson et al. 1994). 
They are usually fitted by a Sunyaev-Titarchuk \shortcite{sun:tit80}
thermal comptonization model with optical depth $\tau_{\rm es}\ga1$\/ and electron
temperature $T_{\rm e}\sim10^9$ K.
These hard spectral components exclude as an interpretation all optically thick accretion models,
like the standard thin accretion disc model of Novikov \& Thorne (1973; hereafter NT73)
or the 'slim accretion disc' model of Abramowicz et al. \shortcite{abr:etal88},
both of which cannot attain temperatures higher than a few $\sim10^7$ K.
The first hot optically thin disc model in agreement with hard spectral components was the
two-temperature accretion disc model by Shapiro, Lightman \& Eardley \shortcite{sha:lig:ear76},
with electron temperature $T_{\rm e}\sim10^8-10^9$ K.
However, this model was found to be thermally unstable due to the temperature-density
dependence of the cooling rate by local bremsstrahlung emission.
A related branch of hot optically thin disc solutions was found,
when radiative cooling channels were extended towards higher temperature domains,
including synchrotron radiation and comptonization of synchrotron and bremsstrahlung photons.
These solutions are thermally and viscously stable for long-wavelength perturbations.
The first models of this class were local thermal equilibrium models,
i.e. the equation of energy balance is solved locally for any radius from a given underlying
disc structure.
Narayan \& Yi (1994, 1995a, 1995b) 
use a self-similar disc structure with sub-Keplerian 
angular momentum distribution (Begelman \& Meier 1982; 
Spruit et al. 1987), whereas Abramowicz et al. \shortcite{abr:etal95} 
and Chen \shortcite{che95} assume a slim disc structure with Keplerian angular momentum distribution.
In spite of these kinematic and other minor differences in the particular cooling functions,
both approaches yield the same local thermal equilibria \cite{che:etal95}.
Advection becomes relevant if temperatures are very high, such that the relative half-thickness 
of the disc, $h/r$\/, is not negligibly small.
This is the case in hot optically thin disc flows,  
where possible heating mechanisms primarily heat the ions to nearly virial temperature.
Since Coulomb collisions become inefficient the electrons gain only little of the thermal
energy of the ions and remain at temperatures of $T_{\rm e}\sim10^9-10^{10}$ K due to inefficient cooling.
The local thermal equilibrium solutions are in general globally advection-dominated.
Morphologically they resemble the ion-supported torus model of Rees et al. \shortcite{ree:etal82}.
It turns out \cite{abr:etal95} that the branch of stable optically thin
equilibrium solutions is limited to a maximum accretion rate $\dM_{\rm max}\sim \alpha^2$\/,
as long as the viscosity parameter $\alpha$\/ is smaller than some critical value
$\alpha_{\rm crit}$\/, depending on the details of the model \cite{che:etal95}.

Recently, global solutions of hot optically thin accretion discs have been calculated. 
This requires to solve the full system of coupled equations of motion for the fluid in the disc,
causing considerably higher numerical efforts due to the sonic point in the radial Euler equation.
Viscous transonic solutions were calculated within the pseudo-Newtonian potential
\cite{pac:wii80} for wedge-shaped flows by Chakrabarti \shortcite{cha96b} and for disc-like flows
by Chen et al. \shortcite{che:etal96} and Narayan et al. \shortcite{nar:etal96}.
It turned out, that for most radii, the local thermal equilibrium solutions yield a good approximation 
for the temperature distribution in the disc.
At the inner edge of the accretion disc, located at the horizon of the black hole, 
matter is forced to corotate with the black hole surface. 
This condition can not be implemented properly within the pseudo-Newtonian potential formalism.
Therefore, disc properties in the very vicinity of the black hole are to be studied in 
full relativity.
Furthermore both, stellar mass black hole candidates in galactic binary systems 
as well as the putative supermassive black holes in the centers of active galactic nuclei
are expected to generally possess a considerable amount of angular momentum.
In the first case the hole is spun up by accretion through the disc,
whereas in the latter case the existence of jets gives 
theoretical support for a rapidly rotating central object 
(e.g. Blandford 1977; Rees et al. 1982). 

Accretion flows around rotating black holes have to be studied in Kerr geometry.
A set of equations for stationary disc-like viscous transonic accretion flows around rotating black holes
has been proposed by Lasota (1994; hereafter L94).
Global transonic solutions of this system, corrected for some misprints, were given by Peitz \shortcite{pei94}
for a polytropic equation of state.
Solutions of a related set of equations, describing hot 
optically thin discs cooled by bremsstrahlung emission,
have recently been calculated by Abramowicz et al. (1996, hereafter ACGL96) and discussed by
Lasota \& Abramowicz (1996).
So far, most formulations for angular momentum transport have been using approximations for the shear, 
which are inconsistent with the kinematics in the supersonic inner part of the accretion discs.

In this paper we derive the system of equations governing stationary viscous transonic disc flows in Kerr geometry
(sections \ref{sec2}, \ref{sec3}).
In particular, we give the consistent expression for the equation of angular momentum balance,
important for the boundary condition at the horizon.
We then discuss in section \ref{sec4} the properties of critical points in inviscid transonic disc-like 
flows, which hold as well for viscous transonic flows in the limit of small viscosity.
In section \ref{sec5} we describe a numerical method to solve the structure equations for transonic flows
and present global solutions for a polytropic equation of state.
%
%
\section{Basic equations}
\label{sec2}
\subsection{Metric}
The half-thickness $h$\/ of the disc is assumed to satisfy the condition $h/r\la1$\/.
Then a power series expansion of the Kerr metric written in cylindrical coordinates 
$\{t,\phi,r,z\}$\/, to zeroth order in the small quantity $z/r$\/, yields a first 
approximation for the underlying disc metric. 
In geometrical units $(G=c=1)$\/, which are used throughout this paper, 
this metric $\vg$\/ reads (NT73)
\beq
\vg
=
-\frac{r^2\Delta}{A}\vd t^2
+\frac{A}{r^2}(\vd\phi-\omega\vd t)^2
+\frac{r^2}{\Delta}\vd r^2
+\vd z^2\;,\non
\label{eq1}
\eeq
where
\beas
\Delta
&=&
r^2-2Mr+a^2\;,\\
A
&=&
r^4+r^2 a^2+2Mra^2\;,\\
\omega
&=&
-g_{t\phi}/g_{\phi\phi}
=
2Mar/A\;.
\eeas
Here $M$\/ is the black hole mass and $a=J/M$\/ its specific angular momentum. 
$\Delta$\/ vanishes at the event horizon, given by $\rh=M+(M^2-a^2)^{1/2}$\/. 
A subscript 'h' refers to quantities evaluated at $\rh$\/.
\subsection{Conservation laws}
Primary variables are the particle current $\vn$\/, 
the stress-energy tensor $\vT$\/ and the entropy flux $\vs$\/. 
The motion of the fluid is governed by conservation of $\vn$\/ and $\vT$\/, 
written as
\bea
0
&=&
\nabla\cdot\vn
\label{eq2}
\;,\\
0
&=&
\vh\cdot(\nabla\cdot\vT)
\label{eq3}
\;,\\
0
&=&
\vu\cdot(\nabla\cdot\vT)
\label{eq4}
\;,
\eea
where $\vu$\/ is the four velocity of the fluid and $\vh=\vu\otimes\vu+\vg$\/ 
projects into its {\it local rest frame} (LRF).
Equations (\ref{eq2}), (\ref{eq3}), (\ref{eq4}) are the continuity equation, 
the Navier-Stokes equations and the energy equation, respectively.

The particle current is $\vn=n\vu$\/ with the baryon number density $n$\/.
The stress-energy tensor for a single component fluid with viscosity and heat flux is
\beq
\vT
=
(\rho+p)\vu\otimes\vu
+p\vg+\vt+\vq\otimes\vu
+\vu\otimes\vq\;.
\label{eq7}
\eeq
Here $\rho=\rho_0+\varepsilon$\/ is the total energy density,
$\rho_0=nm_{\rm B}$\/ is the baryon rest mass density
($m_{\rm B}$\/ being the mean baryon rest mass), $\varepsilon$\/
is the internal energy density and $p$\/ the isotropic pressure.
All thermodynamic quantities refer to the LRF.
Viscous contributions to $\vT$\/ are contained in the viscous stress tensor $\vt$\/, given by
\beq
\vt
=
-\zeta\Theta\vh-2\eta\vsigma\;,\non
\label{eq8}
\eeq
where $\vsigma$\/ is the shear tensor,
the scalar functions $\eta$\/ and $\zeta$\/ are the shear and bulk viscosity coefficients,
respectively, and $\Theta=\nabla\cdot\vu$\/ is the expansion of the fluid world lines.
The energy flux relative to the LRF is described by the spacelike vector $\vq$\/.
The four-entropy flux is defined as
\beq
\vs
=
ns\vu+\vq/T\;,
\label{eq9}
\eeq
with $s$\/ the specific entropy density and $T$\/ the thermodynamic temperature.

Stationarity and axial symmetry of $\vg$\/
imply the existence of two commuting Killing vector fields,
$\vk=\vec{\partial}_t$\/ and
$\vm=\vec{\partial}_\phi$\/,
which give rise to two conserved currents,
\bea
\vP
&=&
\vT\cdot\vk\;,
\;\;\;
\nabla\cdot\vP
=
0\;,
\label{eq5}\\
\vJ
&=&
\vT\cdot\vm\;,
\;\;\;
\nabla\cdot\vJ
=
0
\label{eq6}\;.
\eea
For ideal flows ($\vt=\vq=0$\/), Eqs. (\ref{eq5}), (\ref{eq6}) yield
two constants of motion, namely energy $E=-\mu u_t$\/ and angular momentum $L=\mu u_\phi$\/,
where $u_t$\/ is the binding energy and $\mu=(\rho+p)/\rho_0$\/ the specific 
relativistic enthalpy. 
The specific angular momentum $\lam=L/E=-u_\phi/u_t$\/ is introduced for later use.
\subsection{Velocity field and reference frames}
We consider accretion flows with a four velocity field $\vu=(u^t,\Omega u^t,u^r,0)$\/,
where $\Omega=u^\phi/u^t=d\phi/dt$\/ is the angular velocity and 
$u^r=dr/d\tau$\/ the radial velocity component, negative for accretion.
In addition to the LRF attached to the fluid in motion with $\vu$\/,
two other reference frames are frequently used in relativistic accretion disc theory.

The {\it locally non-rotating frame} (LNRF) is characterized by $u_\phi=0$\/ and orbiting with 
angular velocity $\omega$\/.
Tensor indices with hat '$\;\hat{\;}\;$\/' refer to LNRF.
The physical 3-velocity with respect to LNRF has azimuthal and radial components
\beq
\vph
=\frac{A\tOmega}{r^2\Delta^{1/2}}\quad,\quad
\vrh
=\frac{u^r}{\gam\cU}\;,
\label{eq10}
\eeq
where $\tOmega\equiv\Omega-\omega$\/ and $\cU^2\equiv h^{rr}=(u^r)^2+\Delta/r^2$\/.
Here 
\beq
\gam
=
\left(1-(\vph)^2\right)^{-1/2}
=
\left(1-\frac{A^2\tOmega^2}{r^4\Delta}\right)^{-1/2}
\label{eq11}
\eeq
is the azimuthal Lorentz factor w.r.t. LNRF.
Using $\gam$\/ allows to write the normalization
$\vu\cdot\vu=-1$\/ as
\beq
u^t
=
\frac{\gam A^{1/2}\cU}{\Delta}\;,\;\;\;
u_\phi
=
\frac{\gam A^{3/2}\cU\tOmega}{r^2\Delta}\;.
\label{eq12}
\eeq
In the following, we frequently write $u\equiv u^r$\/ and $l\equiv u_\phi$\/. 
In terms of these variables, $\gam$\/ and $\tOmega$\/ are
\beq
\gam
=
\left(1+\frac{\Delta l^2}{A\cU^2}\right)^{1/2}\;,\;\;\;
\tOmega
=
\frac{r^2\Delta l}{A(A\cU^2+\Delta l^2)^{1/2}}\;.
\label{eq13}
\eeq

The {\it corotating frame} (CRF) is orbiting with the angular velocity $\Omega$\/ of the fluid,
being related to the LNRF by a Lorentz boost with $\gam$\/ in the azimuthal direction.
Tensor indices in CRF are labelled by bar '$\;\bar{\;}\;$\/'. 
The radial component of the physical 3-velocity w.r.t. CRF is $\vrb=\gam\vrh=u/\cU$\/.

The {\it dragging of inertial frames} forces any matter at the horizon to corotate with the surface of
the black hole, $\tOmega_{\rm h}=0$\/ or $\gamh=1$\/, which follows from Eq. (\ref{eq13}).
Consequently, $\vrb_{\rm h}$\/ and $\vrh_{\rm h}$\/ are both equal to the speed of light by virtue of the metric.
This is in contrast to accretion onto non-collapsed objects, where corotation with its
solid surface is imposed as a boundary condition.
For later reference we further introduce the total Lorentz factor w.r.t. LNRF,
\beq
\gamt
=\left(1-(\vph)^2-(\vrh)^2\right)^{-1/2}
=\gam(1-(\vrb)^2)^{-1/2}\;,
\label{eqlamtot}
\eeq
which is singular at the horizon due to $\vrb_{\rm h}=1$\/.
%
%
\section{Equations of disc structure}
\label{sec3}
The flow is treated as a 1-D stationary problem, 
where the disc midplane coincides with the equatorial plane of the hole, 
mass increase of the hole and self-gravity of the disc are neglected. 
We use vertically integrated thermodynamic variables.
Then the equations governing radial and vertical disc structure decouple.
\subsection{Continuity equation}
From the continuity equation $(nu^\alpha)_{;\alpha}=0$\/ it follows that
the total flux of particles through a surface of constant $r$\/ is a constant,
the accretion rate $\dM$\/, given by
\beq
\dot{M}=-2\pi ru\mSigma_0\;,
\label{eq14}
\eeq
where
\beq
\mSigma_0=\int^h_{-h}\rho_0dz\simeq 2h\rho_0
\label{eq15}
\eeq
is the surface rest mass density. For later use we 
introduce two other vertically integrated thermodynamic quantities,
namely the total surface density, $\mSigma\simeq 2h\rho$\/,
and the vertically integrated pressure, $P\simeq 2h p$\/. 
\subsection{Radial momentum equation}
The radial Navier-Stokes equation $h^r_\alpha T^{\alpha\beta}_{;\beta}=0$\/ 
can be written as
\beq
u'u+u^2\cG+\cU^2\cH
=
-(\rho+p)^{-1}(\cU^2p'+S_r+Q_r)\;,
\label{eq16}
\eeq
with $\cG$\/, $\cH$\/, $\cK$\/ defined by 
\bea
\cG&=&(\ln\sqrt{g_{rr}})_{,r}=(a^2-Mr)/(r\Delta)\;,
\label{eq17}\\
\cH&=&\gamma^2 AM\cK/(r^2\Delta)^2\;,
\label{eq18}\\
\cK&=&(\Omega-\Omega^+_{\rm k})(\Omega-\Omega^-_{\rm k})/(\Omega^+_{\rm k}\Omega^-_{\rm k})\;.
\label{eq19}
\eea
The angular velocities
\beq
\Omega^\pm_{\rm k}=\pm\frac{M^{1/2}}{r^{3/2}\pm aM^{1/2}}
\label{eq20}
\eeq
correspond to prograde $(+)$\/ and retrograde $(-)$\/ Keplerian orbits.
Partial derivatives w.r.t. $r$\/ are denoted by a prime.
The quantity $\cK$\/ measures the deviation of the angular velocity 
from the Keplerian value, 
with limiting cases $\cK(\Omega=\Omega^\pm_{\rm k})=0$\/ and $\cK(\Omega=0)=1$\/. 
The quantities $S_r=h_{r\alpha}t^{\alpha\beta}_{;\beta}$\/ and
$Q_r=h_{r\alpha}(q^\alpha u^\beta+q^\beta u^\alpha)_{;\beta}$\/
describe contributions due to viscosity and heat flux. 
In the following we assume these forces to be small compared to pressure forces. 
Then the radial Navier-Stokes equation (\ref{eq16}) reduces 
to the radial Euler equation for a perfect fluid.
With the help of the continuity equation, and after integration over the disc height,
the radial Euler equation can be be written as
\beq
\frac{u'}{u}
=
\frac{\cN}{\cD}\;,\;\;\;
\cN
=
\frac{\cad^2}{r}-\frac{u^2}{\cU^2}\cG-\cH\;,\;\;\;
\cD
=
\frac{u^2}{\cU^2}-\cad^2\;,
\label{eq21}
\eeq
where $\cad$\/ is the adiabatic sound speed.
Eq. (\ref{eq21}) possesses a critical point for $\cD=0$\/, 
identified with a sonic point in the CRF.
Since causality requires $\cad<1$\/, the conditions
$|\vrb_{\rm h}|=1$\/ and $|\vrb|<\cad$\/ at $\rout$\/ suffice to conclude, 
that black hole accretion is transonic \cite{sha:teu83}.
Thus any global solution must satisfy
\beq
\cN(\cD=0)=0
\label{eq22}
\eeq
as a regularity condition.
\subsection{Conservation of angular momentum}
The azimuthal Navier-Stokes equation can be expressed in the form of angular momentum conservation
$J^\alpha_{;\alpha}=0$\/, where the conserved current $J^r$\/ is given by
\beq
J^r
=
(\rho+p)u^ru_\phi+t^r_\phi+q^r u_\phi+q_\phi u^r\;.
\label{eq23}
\eeq
Neglecting the angular momentum associated with the heat flux and using the continuity equation yields
\beq
0
=
\rho_0ru^r(\mu u_\phi)_{,r}
+(rt^r_\phi)_{,r}
\label{eq24}
\;.
\eeq
The vertically integrated angular momentum balance follows after integration over $r$,
\beq
\frac{\dot{M}}{2\pi}(L-L_0)=2hrt^r_\phi\;,
\label{eq25}
\eeq
where $L_0$ is an integration constant.
We did not neglect angular momentum associated with the internal energy.
Therefore Eq. (\ref{eq25}) is the general equation of angular momentum balance
(apart from a contribution due to photons) for any temperature, 
not restricted to flows with $\mu\simeq1$\/ as in NT73 or L94.
\subsection{Description of shear tensor and viscosity}
We assume the coefficient of bulk viscosity to vanish ($\zeta=0$\/). 
The viscous stress tensor then reduces to $\vt=-2\eta\vsigma$\/. 
The shear tensor component $\sigma^r_\phi$\/ can be written as (Eq. (\ref{eqa4}))
\beq
2\sigma^r_\phi
=
\cU^2 l\left(\frac{l'}{l}+\cH\right)
+\frac{u^2 l}{3}\left(\frac{\cN}{\cD}-\frac{2}{r}+3\cG\right)
+2\left[(Ma^2-r^3)\Omega-Ma\right]\frac{\gamma A^{1/2}\cU}{r^4}\;.
\label{eq26}
\eeq
As demonstrated in App. \ref{app1}, expression (\ref{eq26}) may be approximated by
\beq
2\sigma^r_\phi
=
\cU^2 l'
+2\left[(Ma^2-r^3)\Omega-Ma\right]\gamma A^{1/2}\cU r^{-4}\;.
\label{eq27}
\eeq
Based on the dimensional estimate for the viscosity due to the motion of turbulent 
fluid elements in the disc, $\eta\sim\rho_0 v_{\rm turb}l_{\rm turb}$\/, 
we use a parametrization for the kinematic viscosity coefficient $\nu=\eta/\rho_0$\/, 
given by
\beq
\nu
=
\alpha \cad r^\beta f_{\rm c}\;,
\label{eq28}
\eeq
where $\alpha$\/ is the viscosity parameter and $l_{\rm turb}\sim r^\beta$\/
allows to mimic the standard assumption $l_{\rm turb}\sim h$\/
as well as other dependencies, such as $l_{\rm turb}\sim P/P'$\/ 
\cite{pap:sta86}.
Viscosity is a transport process and operates on a finite time-scale, $\tau_\nu$,
limited by causality, no matter whether it is mediated by turbulent gas motion,
magnetic fields, or photons. $\tau_\nu$ becomes equal or larger than
the accretion time-scale $\tau_{\rm acc}$ as the accreted material approaches
the horizon, and as a consequence the viscosity vanishes there,
\beq
\nu_{\rm h}=0\;.
\label{eq29}
\eeq
A cut-off function $f_{\rm c}$\/ has been introduced to guarantee such a behaviour.
We choose \cite{nar92}
\bea
f_{\rm c}
&=&
\left\{
\begin{array}{l@{\quad;\quad}l}
\;\left(1-\left(\vrb/c_\nu\right)^2\right)^2&|\vrb|\le c_\nu\;,\\
\;0&|\vrb|> c_\nu\;,
\end{array}\right.
\label{eq30}
\eea
parametrized by the maximum velocity of viscous information transport, $c_\nu$\/. 
Since $\nu_{\rm h}=0$\/, viscous stresses $t^r_\phi$ vanish at the horizon 
({\it no-torque} condition; Abramowicz \& Prasanna 1990) and the integration constant 
$L_0$ in Eq. (\ref{eq25}) is the angular momentum accreted by the black hole.
Its value is obtained as part of the solution. 

In the limiting case of a Keplerian disc, i.e. for $r\ge\rms$\/ and $u^r\to 0$\/, $u^r_{,r}\to 0$\/, 
$l\to\lk$\/, $\Omega\to\Omegak$\/, Eqs. (\ref{eq26}), (\ref{eq27}) both reduce to the expression of NT73,
\beq
2\sigma^r_\phi
=\gamk^3 A^{3/2}\Delta^{1/2}r^{-5}\Omegak'\;,
\label{eqsigmaNT73}
\eeq
where $\gamk=\gam(\Omegak)$\/ denotes the Keplerian Lorentz factor w.r.t. LNRF. 
For non-Keplerian accretion flows, L94 proposed to replace $\gamk$\/ in Eq. (\ref{eqsigmaNT73})
by the azimuthal Lorentz factor w.r.t. LNRF, $\gam$, as in Eq. (\ref{eq11})
and $\Omegak'$\/ by $\Omega'$\/.
Since $\gamh=1$\/, the resulting expression vanishes at the horizon,
and thus leads to a {\it no-torque} condition as well.
It does so by constraining the kinematics instead of using the physically motivated
condition (\ref{eq29}). 
In addition, this expression leads to a reversal of shear in the vicinity of the black hole, where $\Omega'$\/
changes sign.
Recently, ACGL96 replaced $\gam$\/ in the L94 expression by the total Lorentz factor w.r.t. LNRF, $\gamt$\/, 
as in Eq. (\ref{eqlamtot}). 
Different expressions for $\sigma^r_\phi$\/ are compared in App. \ref{app1}.
\subsection{Conservation of energy}
The energy conservation equation $u^\alpha T^\beta_{\alpha;\beta}=0$\/
can be combined with the continuity equation to 
\beq
\rho_0u^rTs_{,r}
=
-\sigma_{\alpha\beta}t^{\alpha\beta}
-1/3\Theta t^\alpha_\alpha
-q^\alpha a_\alpha
-q^\alpha_{;\alpha}\;,
\label{eq31}
\eeq
where $\va=\nabla_{\vu}\vu$\/ is the four-acceleration of the fluid.
The compressive heating rate $-1/3\Theta t^\alpha_\alpha$\/ vanishes due to the assumption $\zeta=0$\/.
The special relativistic term $\va\cdot\vq$\/ describes the inertia of the heat flux.
Neglecting this term, the vertically integrated energy equation is given by
\beq
Q_{\rm adv}=Q_+-Q_-\;,
\label{eq32}
\eeq
where $Q_{\rm adv}=u\mSigma_0 Ts_{,r}$\/ describes the advection of entropy 
by the radial motion of the fluid, $Q_+=-2h\sigma_{\alpha\beta}t^{\alpha\beta}$\/
is the rate of viscous heat generation and $Q_-=-2hq^z-2hr^{-1}(rq^r)_{,r}$\/
is the local cooling rate.
\subsection{Equation of state}
As a first approximation we assume a polytropic relation between the vertically integrated
pressure $P$\/ and the surface rest mass density $\mSigma_0$\/ \cite{pop:nar91},
\beq
P
=
K\mSigma_0^{\mGamma}\;,
\label{eq33}
\eeq
where $K(s)$\/ is a constant measuring the entropy of the flow and the
polytropic index for a two-dimensional flow, $N=1/(\mGamma-1)$\/, is constant.
The adiabatic sound speed may then be written as $\cad^2=(dP/d\mSigma)_{\rm ad}$\/,
leading to 
\bea
\cad^2
&=&
\left(\frac{dP}{d\mSigma_0}\right)_{\rm ad}
\frac{\mSigma_0}{\mSigma+P}
=
\frac{\mGamma P/\mSigma_0}
{1+N\mGamma P/\mSigma_0}\;.\nonumber
\eea
Using the continuity equation then yields
\beq
\cad^2
=
\left(N+\dcM^{-1/N}(r|u|)^{1/N}\right)^{-1}\;.
\label{eq34}
\eeq
The 'modified accretion rate'
\beq
\dot\cM
=
\mGamma^N K^N(\dot{M}/2\pi)
\label{eq35}
\eeq
consists of a combination of constants $\dM$\/, $K$\/ and $\mGamma$\/ and thus is a constant.
For fixed $\dM$\/ and $\mGamma$\/, $\dcM$\/ measures the entropy in the flow.
For fixed $K$\/ and $\mGamma$\/, $\dcM$\/ is a measure for the accretion rate.
For $\mGamma>1$\/, the specific relativistic enthalpy in terms of $\cad$\/
is $\mu=(1-N\cad^2)^{-1}$\/.
The polytropic equation of state directly couples the temperature to the kinematics of the flow,
and makes the energy equation obsolete.
Instead of two independent parameters $\dM$\/ and $K$\/, 
one is left with a single parameter $\dcM$\/.
\subsection{Vertical disc structure}
The fluid is assumed to be in vertical hydrostatic equilibrium,
i.e. in the LRF the vertical pressure gradient $p_{,z}$\/ balances 
the vertical tidal acceleration towards the midplain of the disc.
A proper treatment of this force balance requires an expansion of the $z$\/-component
of the Navier-Stokes equation to $\cO(z/r)^1$\/.
As shown by Riffert \& Herold \shortcite{rif:her95}, in a Keplerian disc
the hydrostatic equilibrium, obtained from the vertical Euler equation in this way, 
differs considerably from the expression given by NT73.
For general disc flows with $\vu=(u^t,u^\phi,u^r,0)$\/ the vertical Euler equation
\footnote{The derivation of the thickness of transonic accretion discs 
form the vertical Euler equation was proposed by J.P. Lasota, who also drew our attention
towards a different approach by Abramowicz et al. \shortcite{abr:lan:per96}.}
$h^z_\alpha T^{\alpha\beta}_{;\beta}=0$\/, expanded to $\cO(z/r)^1$\/, may be written as (App. \ref{app2})
\bea
\frac{1}{\rho+p}\;\frac{\partial p}{\partial z}
&=&-\frac{z}{r}\left[\frac{M}{r^2}\tcR + \cad^2\cZ\left(\frac{\cN}{\cD}+\frac{1}{r}\right)\right]\;,
\label{neweq3}\\
\tcR
&=&\left[\cY(1-a\Omega)^2-4 a\Omega(1-a\Omega)+2r^2\Omega^2\right]\frac{A\gam^2\cU^2}{\Delta^2}
-\frac{r^2}{\Delta}u^2\;,
\label{neweq4}\\
\cY
&=&1-\frac{2M}{r}+\frac{3a^2}{r^2}\quad,\quad
\cZ
=\frac{2M}{r}-\frac{a^2}{r^2}\;.
\label{neweq5}
\eea
It can be shown that $\tcR$\/ remains finite at $\rh$\/ and therefore (\ref{neweq3})
is regular everywhere. 
In the limiting case of a (cold) Keplerian disc flow,
e.g. for $\cad\to0$\/, $u\to0$\/, $\Omega\to\Omegak$\/, $\rho+p\to\rho_0$\/,
Eq. (\ref{neweq3}) reduces to the formula given by Riffert \& Herold \shortcite{rif:her95}.
To obtain an explicit expression for the half-thickness $h$\/ from Eq. (\ref{neweq3}),
we substitute $\partial p/\partial z\simeq p/h$\/ and assume
\beq
\frac{p}{\rho+p}
\simeq\frac{\mSigma_0}{\mSigma+P}\frac{P}{\mSigma_0}
\simeq\frac{\mSigma_0}{\mSigma+P}\left(\frac{dP}{d\mSigma_0}\right)_{\rm ad}
=\cad^2\;.
\eeq
This leads to
\beq
\frac{h^2}{r^2}
=\frac{\cad^2 r}{M\tcR+\cad^2 r^2\cZ(\cN/\cD+1/r)}\;.
\label{neweq6}
\eeq
If the radial pressure gradient in the vertical Euler equation is negligible, 
(\ref{neweq6}) reduces to $h^2/r^2=(r/M)\cad^2\tcR^{-1}$\/.

The incomplete NT73 expression for the vertical hydrostatic equilibrium in Keplerian flows,
given for comparison with alternative expressions for the disc height, reads
\beq
\frac{1}{\rho_0}\;\frac{\partial p}{\partial z}
=-\frac{z}{r}\frac{M}{r^2}\cR\quad,\quad
\cR
=\gamk^2
\left[\frac{(r^2+a^2)^2+2\Delta a^2}
{(r^2+a^2)^2-\Delta a^2}\right]\;,
\label{neweq2}
\eeq
where $\gamk$\/ is the Keplerian Lorentz factor w.r.t. LNRF.
For non-Keplerian accretion discs, L94 proposed to replace $\gamk$\/ in Eq. (\ref{neweq2}) 
by the azimuthal Lorentz factor (\ref{eq11}).
Recently, ACGL96 used the total Lorentz factor (\ref{eqlamtot}) instead of $\gamk$\/.
Since $\gamt$\/ is singular at the horizon, the disc half-thickness $h$\/,
as derived with this approximation, vanishes at $\rh$\/, which leads 
to infinite densities and therefore to difficulties in the energy equation.
The relative disk height (\ref{neweq6}), as well as alternative expressions, 
are shown in Fig. 12.
Equation (\ref{neweq6}) yields the complete expression as long as the vertical velocity
component $u^z$\/ can be neglected.
%
%
\section{Solution topologies}
\label{sec4}
The regularity condition at the sonic point allows to identify different solution topologies
for accretion flows \cite{abr:zur81}. This analysis, performed for a pseudo-Newtonian
potential, as well as related studies for wedge-shaped flows in Kerr geometry \cite{cha90}
are generalized in this section.
A subscript 's'\/ refers to 'critical quantities', i.e. quantities at $\rs$\/.

As shown in App. \ref{app3}, the regularity condition $\cN(\cD=0)=0$\/ selects a 2-D 
hypersurface $\cF$\/ of the parameter space spanned by three critical variables, 
e.g. by ($\rs$\/, $\cs$\/, $\Omegas$\/) or by ($\rs$\/, $\Es$\/, $\lams$\/).
Projections of $\cF$\/ onto the $\Es$\/-$\lams$\/ plane 
are plotted in Fig. 1 for a non-rotating and for a rotating ($a=0.95$\/) black hole.
%
%
For $\Es$\/, $\lams$\/ within $ABC$\/, three solutions for $\rs$\/ exist.
These are referred to as \inner critical point, $\rsin$\/, \inter critical point, 
$\rsint$\/, and \outer critical point, $\rsout$\/, according to their relative location.
For $\Es$\/, $\lams$\/ beyond the lines $BC$\/ and $BA$\/, only one solution for $\rs$\/ exists,
corresponding to an \inner critical point for $\Es$\/, $\lams$\/ within {\it I} and
to an \outer critical point for $\Es$\/, $\lams$\/ within {\it O}.
\Inner critical points are located in the vicinity of the marginally stable orbit $\rms$\/ 
($\rms=6\,\rg$\/ for $a=0$\/ and $\rms\simeq1.94\,\rg$\/ for $a=0.95$\/, 
$\rg=M$\/ being the gravitational radius),
whereas \outer critical points are located farther away from the black hole (App. \ref{app3}). 
The type of critical points depends on the critical slope $(d\lam/dr)_{\rm s}$\/,
and separates unphysical from possible physical solutions.

In ideal flows, with $\lam=\lams$\/ and $E=\Es$\/ being conserved, $\rsint$\/ is an 
center-type critical point.
Any other critical points are saddle-type, corresponding to possible physical solutions.
Ideal solutions are determined by two out of three constants of motion $E$\/, $\lam$\/, $\dcM$\/.
For given $E$\/ and $\lam$\/, the corresponding accretion rate $\dcM$\/ increases as $\rs$\/ decreases.
For any $\lam$\/ within $ABC$\/, there exists an energy $\Em$\/ such that
the corresponding $\dcM$\/ for a solution passing $\rsin$\/ is equal to $\dcM$\/ for 
a solution passing $\rsout$\/, $\dcMm(\Em,\lam,\rsin)=\dcMm(\Em,\lam,\rsout)$\/
\cite{cha90}. 
Thus the corresponding line $BD$\/ indicates, where stationary solutions actually 'jump' between
different topologies as $E$\/, $\lam$\/ are varied continously.
In the following, solutions with $\Es>\Em$\/ are referred to as \hot,
and solutions with $\Es<\Em$\/ as \cold.
Solutions with $\Es=\Em$\/ pass through both $\rsin$\/ and $\rsout$\/ and are therefore global 
accretion solutions only if a shock occurs within $\rsin$\/ and $\rsout$\/. 
The dashed line in Fig. 1 indicates the maximum $\lam$\/ which a global ideal transonic flow with a 
given $E$\/ may have in order to pass the centrifugal barrier. 

Viscous solutions with non-conserved $E$\/ and $\lam$\/ are represented by curves $E(\lam)$\/. 
The location of $\rs$\/ results from the position of $\Es$\/ and $\lams$\/ on that curve, 
which is generally unknown in advance. 
The solution topology at $\rs$\/, depending on $(d\lam/dr)_{\rm s}$\/, is now determined by the
equation of angular momentum balance, introducing additional parameters, e.g. $\alpha$\/ and $L_0$\/.
For moderate $(d\lam/dr)_{\rm s}$\/, the topological classification in terms of local
critical quantities $\Es$\/ and $\lams$\/ yields essentially the same result as for ideal flows.
The $E(\lam)$\/ curves of such flows may then be superposed on Fig. 1.
If $E(\lam)$\/ never crosses the line $BD$\/, the solution topology and the approximate 
location of $\rs$\/ are known. 
This situation applies e.g. for quasi-ideal flows with globally small gradients 
$dE/dr$\/ and $d\lam/dr$\/, where $E(\lam)$\/ is bound in narrow regions around $\Es$\/ and $\lams$\/. 
If $E(\lam)$\/ crosses the line $BD$\/, the unknown position of $\Es$\/ and $\lams$\/ becomes relevant.
In special cases $\Es$\/ and $\lams$\/ are restricted, e.g. if the flow is assumed ideal within $\rs$\/.

Figure 1 shows that flows around rotating black holes possess generally higher energies and 
thus higher temperatures than flows around non-rotating black holes with corresponding topology.
%
%
\section{Global solutions}
\label{sec5}
\subsection{Model equations and numerical method}
The system of two coupled ODEs solved in this section consists of
Eq. (\ref{eq21}) and Eq. (\ref{eq25}) combined with Eq. (\ref{eq27}),
\bea
\frac{du}{dr}
&=&
\frac{u\cN}{\cD}\;,
\label{eq39}
\\
\frac{dl}{dr}
&=&
\frac{u}{\cU^2}\frac{(L-L_0)}{\nu}
+2\left[Ma+(r^3-Ma^2)\Omega\right]\frac{\gamma A^{1/2}}{\cU r^4}\;.
\label{eq40}
\eea
The adiabatic sound speed $\cad$\/ is calculated according to Eq. (\ref{eq34}),
assuming $N=3/2$\/ ($\mGamma=5/3$\/).
The kinematic viscosity coefficient $\nu$\/ is parametrized by Eq. (\ref{eq28}),
using $c_\nu=1$\/ and $\beta=1$\/.
The inner edge of the disc is located at $\rin=\rh$\/, the outer edge at $\rout=100\rg$\/.
The radial Euler equation has a sonic point at radius $\rs$\/, which is calculated
as part of the solution by the condition $\cD=0$\/.
This provides, together with the regularity condition $\cN(\cD=0)=0$\/,
the boundary condition for Eq. (\ref{eq39}).
At $\rin$\/ the condition $\nuh=0$\/ requires $\Lh=L_0$\/,
which is used as a boundary condition.
The four boundary conditions for the four unknowns $u$\/, $l$\/, $r_s$\/, $L_0$\/ are
therefore given by $\cD=0$\/, $\cN=0$\/, $\Lh=L_0$\/ and e.g. $\lam(\rout)=\lamout$\/.

Eqs. (\ref{eq39}), (\ref{eq40}) are mapped onto the unit interval in $x$\/,
separately for the sub -and supersonic domain. The resulting four equations are then solved
with $x$\/ as an independent variable. The sonic point is mapped onto $x=0$\/, 
where $\cN=0$\/, $\cD=0$\/ are taken as boundary conditions both for the supersonic and
subsonic equations. 
The horizon and the outer radius of the disc are both mapped onto $x=1$\/,
where $L=L_0$\/ is used as boundary condition for the supersonic equations.
For a given $L_0$\/, the solution is then uniquely determined, including $\lamout$\/.
An iteration scheme in $L_0$\/ is set up to find global solutions for a given $\lamout$\/.
The boundary value problem is integrated by a collocation method \cite{bad:kun89}, 
using a fully adaptive grid with variable number of subintervals. 
This allows to obtain the solution to a pre-specified accuracy. 
A parameter continuation is used to study the variation in $a$\/, $\alpha$\/ and $\dcM$\/,
as well as for the iteration in $L_0$\/.

We use dimensionless quantities by normalizing distances to the gravitational radius $\rg$\/
and velocities to the speed of light (geometrical units). 
For a polytropic equation of state the resulting system is independent of mass $M$\/.
We present ideal and viscous solutions for any of the topologies discussed in Sect. \ref{sec4}.
The relative half-thickness $h/r$\/ of the disc is calculated from Eq. (\ref{neweq6}).
No retrograde disc flows with $\lam<0$\/ are considered.
\subsection{Ideal solutions}
To give examples for all solution topologies, we plot model sequences of increasing $\lam$\/ for 
$E=1.015$\/ in the case of a non-rotating ($a=0$\/) black hole (see Fig. 2) 
and for $E=1.06$\/ in the case of a rotating ($a=0.95$\/) black hole (see Fig. 3). 
The various models correspond to crosses $\times$\/ in Fig. 1.
%
%
For lower $\lam$\/, within $ABD$\/ or {\it O}, solutions pass an {\it outer} critical point.
For $\lam$\/ within {\it O}, $u$\/ grows with decreasing $r$\/.
For $\lam$\/ within $ABD$\/, the existence of the {\it inner} and {\it inter} sonic point
becomes 'visible' in a deformation of $u$\/ at the corresponding locations.
As $\lam$\/ approaches the line $BD$\/ from within $ABD$\/, $u$\/ passes the {\it outer} sonic point
followed by a local maximum and a local minimum, both within the supersonic region. 
This behavior occurs because $\cN=0$\/ is encountered in total three times, 
first together with $\cD=0$\/ at the sonic point,
and thereafter two more times for finite $\cD$\/.
As $\lam$\/ crosses the line $BD$\/, stationary solutions 'jump' from \cold to \hot topology,
passing the {\it inner} critical point (compare e.g. Chakrabarti 1990; 
Kafatos \& Yang 1994). 
As $\lam$\/ approaches the line $BD$\/ from within $BCD$\/, $u$\/ first passes a local maximum
followed by a local minimum, both within the subsonic region, before passing the {\it inner} sonic point.
The majority of \hot solutions with $\lam$\/ within {\it I} have again monotonic $u$\/-profiles.

The 'jump' of stationary solutions at the line $BD$\/ is not visible in the corresponding $\Omega$\/-profiles.
In the vicinity of the horizon, the $\Omega$\/-distribution shows a kinematic boundary layer structure,
dropping with a steep gradient to $\tOmega_{\rm h}=0$\/. 
In that kinematic boundary layer, the maximum of $\Omega$\/ and the steepness of $d\Omega/dr$\/ depend on
the angular momentum of the flow and the rotation of the hole.
This is illustrated in Fig. 4, where $\Omega$\/ is plotted for 
a sequence of \hot models with increasing $a$\/ and fixed $E=1.2$\/, $\lam=1.0$\/.
%
%
%
\subsection{Viscous solutions}
For given boundary conditions the global solution is uniquely determined
and possesses either \hot or \cold topology.
\Hot solutions are generally found for moderate $\alpha$\/ and higher $\dcM$\/.
In Fig. 5 we plot \hot solutions around a non-rotating black hole,
which differ only in $\alpha$\/, with otherwise fixed $a$\/, $\dcM$\/ and $L_0$\/.
They show a globally sub-Keplerian angular momentum distribution. 
%
%
With increasing $\alpha$\/ the angular momentum transport becomes more efficient,
such that practically all reasonable boundary conditions $\lamout$\/,
from $\lamout\ga\lams$\/ to $\lamout\la\lamk(\rout)$\/ are met.
For $\alpha$\/ fixed, $\lamout$\/ increases with $L_0$\/.
This is demonstrated in Fig. 6 for a sequence of \hot solutions around a
rotating ($a=0.95$\/) black hole,
which are for the same $a$\/, $\alpha$\/ and $\dcM$\/, but differ in $L_0$\/.
%
%
There exist solutions which approach a Keplerian distribution $\lamk$\/ globally or even exceed 
$\lamk$\/ locally.
These require high $\lam$\/ in the vicinity of $\rms$\/, obtained for lower $\alpha$\/ and possess higher $L_0$\/.
In Fig. 7 we plot two examples of such trans-Keplerian solutions for $a=0$\/.
Solutions of that kind possess extremely high temperature at $\rout$\/, such that their
physical relevance appears doubtful.
%
%
For very low viscosity, e.g. for $\alpha=10^{-4}-10^{-3}$\/, accretion profiles are quasi-ideal.
Under these conditions, \hot solutions with $E(\lam)$\/ within $BCD$\/ can be calculated,
possibly passing local extrema in their $u$\/-profiles. 
In particular, the quasi-constant $\lam$\/ may exceed $\lamk(\rms)$\/ in the vicinity of $\rms$\/.
In Fig. 8 we plot examples of \hot quasi-ideal solutions for $a=0$\/,
with globally sub-Keplerian as well as locally super-Keplerian $\lam$\/-distribution.
For higher $\lam$\/ or for energies below unity, $\rsout$\/ is shifted outwards,
in extreme cases beyond $\rout$\/ (Fig. 14).
Then even solutions with inverted angular momentum transport in their sub-Keplerian 
outer parts exist (Fig. 8), as reported by Abramowicz et al. \shortcite{abr:etal88}.
However, in order to guarantee that these are global solutions, 
the integration must be extended to larger $\rout$\/.
%
%

Except from quantitative differences, \hot solutions show the same qualitative
behavior for any $a$\/.
The $\cad$\/-distribution and thus the temperature distribution possess a maximum outside $\rs$\/.
For given $a$\/, $\dcM$\/ and $L_0$\/, higher $\alpha$\/ yield steeper $\lam$\/-profiles and 
higher temperatures.
Therefore, quasi-ideal flows represent the limit of lowest temperature discs for a set of 
$a$\/, $\dcM$\/ and $L_0$\/.
Since temperature remains high at $\rout$\/, no (cold) standard thin disc configuration is met.
This reproduces a well-known property of hot accretion flows,
and is not an artefact of a polytropic equation of state.
A common feature of \hot solutions is a very flat $\lam$\/-distribution inside $\rs$\/.
This is due to low $\nu$\/ resulting from the decrease of $\cad$\/ within $\rs$\/.
The energy in the disc increases moderately towards $\rout$\/. 
The decrease of $\cad$\/ towards $\rh$\/ is imprinted on the $E$\/- profiles and the 
$L$\/-profiles (not plotted), which show steeper gradients than $\lam$\/ in the vicinity of $\rh$\/.
Thus, in general, even for quasi-flat $\lam$\/-distribution inside $\rs$\/,
the angular momentum $L_0$\/ deposited onto the black hole in a \hot flow is
significantly lower than $L_{\rm s}$\/.
The relative disc height respects $h/r\la1$\/ at $\rs$\/, but in models with effective 
angular momentum transport exceeds unity beyond $\simeq20\rg$\/ due to very high temperaturs.
The $\Omega$\/-distributions show a kinematic boundary layer structure similar
to ideal flows as a consequence of flat $\lam$\/-distributions within $\rs$\/,
such that disc models with $\nu=0$\/ in the supersonic domain \cite{pop:nar92}, 
essentially yield similar profiles.

Viscous solutions passing the \outer sonic point exist for higher $\alpha$\/ and lower $\dcM$\/.
Since the angular momentum at $\rsout$\/ is strongly sub-Keplerian, 
\cold solutions possess smaller $L_0$\/.
In Fig. 9 we plot a sequence of \cold solutions around a non-rotating black hole,
for fixed $\dcM$\/ and with different $L_0$\/.
Since $\rs\gg\rh$\/, global solutions with any reasonable boundary condition $\lamout$\/ 
do not exist for each $\alpha$\/.
The two solutions with higher $\lam$\/ in Fig. 9 are for the same $\alpha=0.06$\/,
choosen such that the corresponding maximum possible $\lamout$\/ is Keplerian.
Lower $\alpha$\/ only allow for $\lamout<\lamk(\rout)$\/ (for otherwise fixed parameters).
The two solutions with lower $\lam$\/ in Fig. 9 are for $\alpha=0.05$\/.
The global $\lam=0$\/ distribution with $L_0=0$\/ does no more depend on $\alpha$\/,
describing the limit of 'spherical disc accretion' around a non-rotating hole.
This limit is reproduced for any $E$\/ and $\lam$\/ within Fig. 1.
%
%
Alternatively, as for \hot solutions, different $\lamout$\/ can be found for 
fixed $a$\/, $\dcM$\/ and $L_0$\/ by variation of $\alpha$\/. 
This requires $\dcM$\/ high enough in order to achieve $\rs<\rout$\/.
A sequence of such \cold solutions around a rotating ($a=0.95$\/) black hole is shown in Fig. 10.
%
%
As for \hot flows, quasi-ideal solutions with $E(\lam)$\/ within $ABD$\/ exist, 
again similar to corresponding ideal solutions and not shown here.

Solutions around rotating and non-rotating black holes differ only quanitatively.
Since the $\lam$\/-distribution is strongly sub-Keplerian at $\rs$\/,
accretion flows with smaller $\lamout$\/ result in general and no \cold solution with a global
$\lamk$\/-distribution does exist.
Nevertheless, $\lam$\/ can be steeper within $\rs$\/ than for \hot solutions due to higher $\alpha$\/.
\Cold solutions have generally lower temperature than \hot, with a maximum at the horizon.
However, as before, no (cold) standard disc configuration is matched at $\rout$\/.
With decreasing $\dcM$\/, $\rs$\/ moves outward and $\cad$\/ decreases
due to lower $\lams$\/ and thus less efficient global angular momentum transport. 
Therefore the relative disc height $h/r$\/ exceeds unity only in the very outer disc regions.
The kinematic boundary layer is less prominent due to generally lower $\lam$\/. 
We do not find solutions with negative accreted angular momentum.
%
%
\section{Summary and discussion}
\label{sec6}
We derived the equations of motion for stationary viscous disc-like accretion 
flows around rotating black holes.
In particular, we give a complete expression for the shear tensor that governs
the transport of angular momentum in relativistic transonic disc flows.
In order not to violate causality, the kinematic viscosity has to vanish at the horizon.
This is achieved by an appropriate parametrization.
Consequently, no torque is exerted on the horizon.
It is shown that the expression widely used for the shear, 
calculated by assuming circular trajectories, is incompatible with the kinematics in the
supersonic inner part of the accretion discs.
We wish to stress that it is important to use the correct boundary conditions
at the horizon, even if it turns out that they do not strongly affect the global
properties of solutions. The range of parameters, for which global solutions exist,
can be affected by the choice of boundary conditions.
Furthermore, we calculate an expression for the half-thickness of the disc,
obtained by integration of the vertical Euler equation over disc height.

The properties of the sonic point are exploited to classify different solution topologies of
ideal disc-like accretion flows around rotating black holes.
Depending on angular momentum and energy, global solutions pass through the
inner or outer critical point, referred to as \hot and \cold, respectively.
We use a polytropic equation of state for vertically integrated pressure and density to calculate
global viscous solutions. 
The coupled disc equations are solved between the horizon of the black hole and $100\rg$\/.
A numerical method is presented that allows to integrate the structure equations of transonic
accretion flows.
In cases of moderate angular momentum transport, the topological classification for ideal flows
also holds for viscous flows.
We give examples of each topology for non-rotating and rapidly rotating black holes ($a=0.95$\/).
We present solutions, where the angular momentum at the outer edge of the disc varies between
zero and the Keplerian value, for a wide range of parameters.
The angular momentum $L_0$\/ accreted by the black hole is calculated as part of the solution.
For small values of $\alpha$\/ and higher $\dcM$\/, solutions are of \hot, 
with the sonic point located in the vicinity of the marginally stable orbit. 
In these cases, the temperature distribution in the disc has a maximum outside the sonic point.
There exist solutions which are sub-Keplerian everywhere (Figs. 5, 6),
and accretion flows, which are super-Keplerian in some region inside and outside the
marginally stable orbit (Fig. 7, 8).
In this case, the angular momentum accreted by the black hole generally exceeds the
Keplerian value at the marginally stable orbit.
For these partially super-Keplerian flows the sonic point is located between the
photon orbit and the marginally stable orbit.
For a given angular momentum at the outer edge of the disc, trans-Keplerian flows
possess a higher temperature than sub-Keplerian flows.
For higher values of $\alpha$\/ and smaller $\dcM$\/, solutions are of \cold,
with the sonic point located further away from the black hole.
They generally possess a sub-Keplerian angular momentum distribution for any radius.
These solutions are colder than \hot and have their maximum temperature at the horizon.
In the outermost part of \hot solutions, the relative disc height $h/r$\/ exceeds unity, 
or the temperature reaches the virial temperature. 
Our solutions however remain consistent for the inner $20\rg$\/.

We obtain solutions around rotating black holes for any Kerr parameter $a$\/.
Discs around rotating black holes are hotter and deposit less angular momentum
onto the central object than discs around non-rotating black holes.
For the latter the temperature maximum is broader and at larger radii than in the
case of a rapidly rotating black hole.
Temperatures of our solutions lie in the range of $10^{11}-10^{12}$ K. 
They remain high throughout the whole disc and our solutions do not match a (cold) 
standard thin disc configuration at the outer edge.
These results suggest to compare our polytropic model with other hot accretion disc models,
e.g. with optically thin advective disc models.
Global solutions of this type, described by a related system of equations,
have been calculated recently \cite{abr:etal96}. 
It turns out that the general features of both models argee.
A polytropic equation of state therefore provides a reasonable description for
many properties of hot optically thin advective disc flows.
Since we do not find any solution with considerably lower temperature than
$\sim10^{11}$ K, a polytropic equation of state does not seem to reproduce the class of 
cold radiation-pressure supported optically thick slim discs 
\cite{abr:etal88}.
%

We are grateful to M. Camenzind (who initiated this work) and to R. Khanna for 
invaluable discussions.
Stimulating comments by S. Chakrabarti, J. Ferreira, J. P. Lasota and L. Rezzolla are acknowledged.
This work has been partially supported by Deutsche Forschungsgemeinschaft (SFB 328)
and the French Ministery of Foreign Affairs.
\appendix
%
%
\section{Shear tensor component $\sigma^r_\phi$\/}
\label{app1}
To calculate the shear tensor $\vsigma$\/ it is convenient to start from
the decomposition of the gradient of the four-velocity, $\nabla\vu$\/,
into irreducible tensorial parts,
\beq
u_{\alpha;\beta}
=
\sigma_{\alpha\beta}+\omega_{\alpha\beta}
+\frac{1}{3}\Theta h_{\alpha\beta}-a_\alpha u_\beta\;,
\label{eqa1}
\eeq
where $\vec\omega$\/ is the antisymmetric rotation 2-form,
$\vsigma$\/ is the tracefree symmetric shear tensor and
$\va$\/ is the four-acceleration of the fluid.
The expansion of the fluid world lines is given by
$\Theta=u^\alpha_{;\alpha}=u^r_{,r}+u^r/r$\/.
Symmetrization (expressed by brackets) of Eq. (\ref{eqa1}) leads to
\beq
\sigma_{\alpha\beta}
=
u_{(\alpha;\beta)}+a_{(\alpha}u_{\beta)}-\frac{1}{3}\Theta h_{\alpha\beta}\;.
\label{eqa2}
\eeq
The relevant component for the angular momentum equation is then
\beq
2\sigma^r_\phi
=
u^r_{;\phi}
+g^{rr}u_{\phi;r}+a^r u_\phi+a_\phi u^r-\frac{2}{3}\Theta u^r u_\phi\;,
\label{eqa3}
\eeq
which, for $\vu=(u^t,\Omega u^t,u^r,0)$\/, can be written as
\beq
2\sigma^r_\phi
=
\cU^2 l\left(\frac{l'}{l}+\cH\right)
+\frac{u^2 l}{3}\left(\frac{u'}{u}-\frac{2}{r}+3\cG\right)
+2\left[(Ma^2-r^3)\Omega-Ma\right]\frac{\gamma A^{1/2}\cU}{r^4}\;.
\label{eqa4}
\eeq
Instead of using this full expression (with $u'/u=\cN/\cD$\/),
$\sigma^r_\phi$\/ may be approximated by a simplification of the
velocity field $\vu$\/ involved in the calculation of $\sigma^r_\phi$\/.
The simplest approximation of a Keplerian disc flow with $u^r=0$\/ leads to
Eq. (\ref{eqsigmaNT73}). 
While consistent for nearly circular trajectories given in Keplerian discs (NT73),
Eq. (\ref{eqsigmaNT73}) breaks down in the inner part of transonic discs, 
where radial and azimuthal velocities become comparable.

A more appropriate simplification of Eq. (\ref{eqa4}) consists of dropping terms which contain $u'/u$\/.
We therefore neglect the expansion ($\Theta=0$\/) and the radial four-acceleration ($a^r=0$\/).
This approximation yields
\beq
2\sigma^r_\phi
=
\cU^2 l'
+2\left[(Ma^2-r^3)\Omega-Ma\right]\gamma A^{1/2}\cU r^{-4}\;.
\label{eqa6}
\eeq
We found that viscous solutions calculated using Eq. (\ref{eqa6}) are consistent 
with the assumptions made above for all parameters and for almost any radius.
In Fig. 11 the exact expression (\ref{eqa4}) was used to
calculate a typical \cold solution. 
%
%
This solution was used to calculate the approximation (\ref{eqa6}), as well as
other expressions proposed for the shear.
The exact expression (\ref{eqa4}) is plotted as a solid line and remains finite at the horizon.
The approximation (\ref{eqa6}) is plotted as a short dashed line and yields a good approximation 
to (\ref{eqa4}) for all radii. 
Eq. (\ref{eqsigmaNT73}) with $\gam$\/ and $\Omega'$\/ instead of $\gamk$\/ and $\Omegak'$\/ 
(long dashed line; L94) differs considerably, especially in the vicinity of the kinematic boundary layer.
Eq. (\ref{eqsigmaNT73}) with $\gamt$\/ and $\Omega'$\/ instead of $\gamk$\/ and $\Omegak'$\/ 
(short dash-long dashed line; ACGL96) remains a good approximation for larger radii, 
but tends to extremely high values of inversed shear in the inner parts
of the flow, due to the singular behaviour of $\gamt$\/ at $\rh$\/.
For larger radii all expressions approach the Newtonian limit.
%
%
\section{Vertical Euler equation}
\label{app2}
In this section the vertical Euler equation $h^z_\alpha T^{\alpha\beta}_{;\beta}=0$\/ 
is derived for a velocity field $\vu=(u^t,u^\phi,u^r,0)$\/, to an accuracy of first 
order in $(z/r)$\/.
This requires an expansion of the Kerr metric to $\cO(z/r)^2$\/ 
\cite{rif:her95}, denoted by $\vG$\/, and written as
\beq
\vG
=\vg+\frac{z}{r}\;\vgt+\frac{z^2}{r^2}\;\vgtt\;,
\label{eqb1}
\eeq
where $\vg$\/ is the zeroth order (Eq. (\ref{eq1}); NT73), with non-vanishing components
\bea
g_{tt}
&=&\frac{A\omega^2}{r^2}-\frac{\Delta r^2}{A}
=-\left(1-\frac{2M}{r}\right)\quad,\quad
g_{t\phi}
=g_{\phi t}=-\frac{A\omega}{r^2}
=-\frac{2Ma}{r}\;\non\\
g_{\phi\phi}
&=&\frac{A}{r^2}\quad,\quad
g_{rr}
=\frac{\Delta}{r^2}\quad,\quad
g_{zz}
=1\;,
\eea
$\vgt$\/ is the first order correction, with non-vanishing components
\beq
\tg_{rz}=\tg_{zr}=g_{rr}\left(\frac{2M}{r}-\frac{a^2}{r^2}\right)\;,
\label{eqb2}
\eeq
and $\vgtt$\/ is the second order correction, with non-vanishing components
\bea
\ttg_{tt}
&=&-\frac{M}{r}\left(1+\frac{2a^2}{r^2}\right)\;,\non\\
\ttg_{t\phi}
&=&
\ttg_{\phi t}
=-g_{t\phi}\left(\frac{3}{2}+\frac{a^2}{r^2}\right)\;,\non\\
\ttg_{\phi\phi}
&=&-a^2\left(1+\frac{5M}{r}+\frac{2Ma^2}{r^3}\right)\;,\non\\
\ttg_{rr}
&=&-g_{rr}^2\left(\frac{3M}{r}-\frac{4M^2}{r^2}
-\frac{a^2}{r^2}\left(3-\frac{6M}{r}+\frac{2a^2}{r^2}\right)\right)\;,\non\\
\ttg_{zz}
&=&g_{rr}\left(\frac{2M}{r}-\frac{2Ma^2}{r^3}+\frac{a^4}{r^4}\right)\;.
\label{eqb3}
\eea
Neglecting contributions of viscosity and heat in $\vT$\/,
the leading terms of $h^z_\alpha T^{\alpha\beta}_{;\beta}=0$\/ are
\beq
G^{\alpha z}p_{;\alpha}+(\rho+p)u^\alpha u^z_{;\alpha}=0\;.
\label{eqb4}
\eeq
The semicolon denotes covariant derivatives w.r.t. $\vG$\/. This yields
\beq
G^{rz}p_{,r}+p_{,z}
=
-(\rho+p)\left[\Gamma^z_{tt}(u^t)^2+\Gamma^z_{t\phi}u^t u^\phi+\Gamma^z_{\phi\phi}(u^\phi)^2
+\Gamma^z_{rr}(u^r)^2\right]\;,
\label{eqb5}
\eeq
where we already used $G^{zz}=1+\cO(z/r)^2$\/ and $\Gamma^z_{tr}=\Gamma^z_{\phi r}=0$\/.
The remaining expressions in Eq. (\ref{eqb5}) are
\bea
2\Gamma^z_{\alpha\beta}
&=&-G_{\alpha\beta,z}-G^{rz}G_{\alpha\beta,r}\quad,\quad\alpha,\beta\in\{t,\phi\}\;,\\
2\Gamma^z_{rr}
&=&2G_{rz,r}-G_{rr,z}+G^{rz}G_{rr,r}\;,\\
G^{rz}
&=&-(z/r)\cZ+\cO(z/r)^2\;,
\label{eqb6}
\eea
which leads to 
\bea
\Gamma^z_{tt}
&=&
\frac{z}{r}\frac{M}{r^2}\cY+\cO(z/r)^2\;,\\
\Gamma^z_{t\phi}
&=&
-\frac{z}{r}\frac{M}{r^2}a(2+\cY)+\cO(z/r)^2\;,\\
\Gamma^z_{\phi\phi}
&=&\frac{z}{r}\frac{M}{r^2}\left[a^2(4+\cY)+2r^2\right]+\cO(z/r)^2\;,
\label{eqb7}
\eea
where $\cY$\/, $\cZ$\/ are given in Eq. (\ref{neweq5}). 
Equation (\ref{eqb5}) is further reduced by $u^\phi=\Omega u^t$\/ with $u^t$\/ as in Eq. (\ref{eq12})
and the radial pressure gradient is, as in the case of the radial Euler equation (\ref{eq21}) 
expressed in terms of $\cad$\/,
\beq
\frac{1}{\rho+p}\frac{dp}{dr}
\simeq
-\cad^2\left(\frac{u'}{u}+\frac{1}{r}\right)\;.
\eeq
This then leads to the final expression (\ref{neweq3}).
The limiting value of $\cR$\/ at the horizon, $\tcRh$\/,
\beq
\tcR_{\rm h}
=\frac{3}{4M^2}\left[(\lh-a\uh\cZh^{1/2})^2-\lh^2(1-\cZh)\right]+\cZh\;,
\label{eqb8}
\eeq
remains finite.

In Fig. 12 we plot three different expressions proposed for the 
relative half-thickness $h/r$\/, calculated for a typical \cold solution using
the exact expression for the shear (same solution as in Fig. 11).
%
%
The solid line corresponds to Eq. (\ref{neweq6}), which is to be compared with the
corresponding expressions derived from Eq. (\ref{neweq2}). These are 
given by $(h/r)^2=(r/M)\cad^2/\cR$\/, with $\cR$\/ calculated using either $\gam$\/ instead of $\gamk$\/
(short dashed line; L94) or $\gamt$\/ instead of $\gamk$\/ (long dashed line; ACGL96).
All expressions yield similar flow morphology in the outer subsonic parts of the flow.
The disc height (\ref{neweq6}) is finite at $\rh$\/, in agreement with (\ref{eqb8}).
Using $\gam$\/ in (\ref{neweq2}) overestimates $h$\/ within $\sim10\rg$\/,
whereas $\gamt$\/ yields an underestimate, with vanishing $h$\/ at the horizon.
%
%
\section{Existence and location of sonic points}
\label{app3}
In the Kerr metric circular equatorial geodesic orbits
exist outside of the photon orbit $\rph^\pm=2M(1+\cos(2/3\cos^{-1}(\pm a/M))$\/.
At $\rph$\/ the Keplerian angular velocities $\Omegak^\pm$\/ 
reach the limiting angular velocities $\Omega_\pm$\/ for timelike trajectories
(dashed lines in Fig. 13, bottom).
Circular geodesic orbits are unstable within the marginally stable orbit $\rms>\rph$\/.
In the following we use abreviations $\lamph=\lamk(\rph)$\/ and $\lamms=\lamk(\rms)$\/.

For given $\rs$\/, $\cs$\/ the regularity condition $\cN(\cD=0)=0$\/ yields a quadratic in $\Omegas$\/,
\bea
0
&=&
a_0+a_1\Omegas+a_2\Omegas^2\;,
\label{eqc1}\\
a_0
&=&
\cs^2\rs^2(M-\rs)(4M^2a^2-\rs^2\Delta)-MA\Delta
\;,\non\\
a_1
&=&
2MaA[\Delta-2\rs\cs^2(M-\rs)]
\;,\non\\
a_2
&=&
A[\Delta(\rs^3-Ma^2)+A\cs^2(M-\rs)]
\;.\non
\label{eqa47}
\eea
In the limiting case $\cs=0$\/, solutions of Eq. (\ref{eqc1}) reduce to
$\Omegas^\pm=\Omegak^\pm$\/, where $\Omegas^+\ge\omega$\/ corresponds
to prograde and $\Omegas^-<\omega$\/ to retrograde critical points.
In the general case $0<\cs^2<1/3$\/ one finds $\Omegak^+>\Omegas^+$\/
for $\rs^+>\rph^+$\/ and $\Omegak^-<\Omegas^-$\/ for $\rs^->\rph^-$\/.
Therefore transonic disc flows have sub-Keplerian $\Omega$\/-and $\lam$\/-distributions
in the vicinity of the sonic point \cite{abr:zur81}.
A minimum critical radius $r_{{\rm s}-}(\cs)$\/ exists,
which is located between $r_{{\rm s}-}=\rph^\pm$\/ for hot flows with $\cs^2=1/3$\/
and $r_{{\rm s}-}=\rms^\pm$\/ for cold flows with $\cs=0$\/.
The corresponding $\lam$\/ represents an upper limit $\lam_{\rm top}$\/
allowed for an ideal transonic polytropic flow with a given enthalpy 
\cite{kaf:yan94}. 
On the other hand, a maximum critical radius $r_{{\rm s}^+}(\cs)$\/ exists,
which is smallest for hot flows and tends to infinity as $\cs$\/ vanishes. 
In Fig. 13 we plot $\Omegas^\pm(\rs)$\/ and $\lams^\pm(\rs)$\/ 
for different values of $\cs^2$\/ as a parameter,
both for a non-rotating and a rotating ($a=0.95$\/) black hole.
%
%
A limiting critical sound speed $\cs^\ast(a)$\/ separates the topology 
of the $\lams(\rs)$\/ curves in two classes.
If $\cs>\cs^\ast$\/, $\lams(\rs)$\/ is monotonic, 
thus any $\lams\le\lamph$\/ corresponds to one solution for $\rs$\/ and 
$\lams>\lamph$\/ yields no solution at all.
On the other hand, if $0<\cs<\cs^\ast$\/, $\lams(\rs)$\/ has a minimum $\lam_{\rm min}$\/ 
and a maximum $\lam_{\rm max}$\/, which fall together for $\cs=\cs^\ast$\/.
Obviously, $\lam_{\rm min}\le\lamms$\/ and $\lam_{\rm max}\ge\lamms$\/.
Consider first the case $\lam_{\rm max}\le\lamph$\/.
Then for $\lams\le\lam_{\rm min}$\/ again exactly one $\rs$\/ exists, 
for $\lam_{\rm min}<\lams<\lam_{\rm max}$\/ there exist three solutions for $\rs$\/,
and for $\lams\ge\lam_{\rm max}$\/ again one $\rs$\/ exists as long as $\lams\le\lamph$\/,
and no solution otherwise.
The number of critical points in each of these three domains is reduced by one
if $\lam_{\rm max}>\lamph$\/.
These critical points can be identified by the corresponding slope $d\lam_{\rm s}/dr_{\rm s}$\/
and their relative location. 
An \outer critical point $\rsout$\/ exists under multiple critical
point conditions and has $d\lam_{\rm s}/dr_{\rm s}<0$\/.
An \inter critical point $\rsint$\/ has $d\lam_{\rm s}/dr_{\rm s}>0$\/. 
An \inner critical point $\rsin$\/ has $d\lam_{\rm s}/dr_{\rm s}<0$\/ and is
located next to the horizon.
The simple relation (\ref{eqa47}) among $\rs$\/, $\cs$\/, $\Omegas^\pm$\/ is locally relevant only.
This changes for ideal flows if $\cN(\cD=0)=0$\/ is solved for conserved quantities,
since then the solution topology is known from the boundary conditions.
Out of numerous combinations we plot $\lams^\pm(\rs)$\/ parametrized by $\Es$\/ 
in Fig. 14 and $\Es(\rs)$\/ parametrized by $\lams$\/ in Fig. 15,
both for $a=0$\/ and for $a=0.95$\/.
%
%
Again, the same classes of critical points are recovered from both diagrams.
For $\Es<\Es^\ast$\/ there exist extreme values $\lam_{\rm min}$\/ and $\lam_{\rm max}$\/, 
such that for $\lam_{\rm min}<\lams<\lam_{\rm max}$\/ multiple critical points occur. 
Formulated vice versa, for $\lams\ge\lams^\ast$\/ there exist extreme values 
$\Emin$\/ and $\Emax$\/, such that for $\Emin<\Es<\Emax$\/ multiple critical points occur.
This is summerized in Fig. 1, where $\Emin(\lams)$\/ and $\Emax(\lams)$\/ are plotted.

Consider now ideal flows with $\lam$\/ and $E$\/ in the region of three critical points,
and ask through which of these a particular transonic solution passes. 
The answer depends partly on the nature of critical points,
which results from the critical slope $(du/dr)_{\rm s}$\/,
calculated by de l'H\^{o}pital's rule from critical quantities.
It turns out, that $\rsint$\/ corresponds to an unphysical 
center-type critical point (spiral-type in viscous flows), 
whereas the \inner and \outer critical points are saddle-type, 
allowing for physical solutions.
Thus, as $\lam$\/ increases continuously, the stationary solution 'jumps' 
from passing $\rsout$\/ towards passing $\rsin$\/ somewhere between
$\lam_{\rm min}$\/ and $\lam_{\rm max}$\/.
Where this transition occurs in the $E$\/-$\lam$\/-plane, 
depends on the modified accretion rate $\dcM$\/ corresponding to a particular $\rs$\/.
$\dcM$\/ is generally different for solutions passing $\rsin$\/ or $\rsout$\/,
$\dcM(E,\lam,\rsin)\neq\dcM(E,\lam,\rsout)$\/.
However, there exists an energy $\Em$\/, such that 
$\dcMm(\Em,\lam,\rsin)=\dcMm(\Em,\lam,\rsout)$\/. 
The line $BD$\/ in Fig. 1 represents the locus of these $\Em$\/.
%

%
%
\newpage
\section{Figures}
{\bf Figure 1:}
Division of the parameter space spanned by $\Es$\/, $\lams$\/
according to different types of ideal flow, for black holes with $a=0$\/
(top diagram) and $a=0.95$\/.
Curve $ABC$\/ is the boundary of the parameter space which seperates the region 
(between $AB$\/ and $CD$\/) with multiple critical points from the region with a 
single critical point (rest of diagram).
$BD$\/ is the locus of flows for which both, $\Es$\/ and $\dcM$\/,
at the inner and at the outer critical points are the same. 
The dashed line indicates the maximum $\lam$\/ which an ideal flow with a 
given $E$\/ may have in order to pass the centrifugal barrier.
The locations of ideal models presented in section \ref{sec5} are indicated by $\times$\/.

\vspace{1cm}
{\bf Figure 2:}
Sequence of ideal disc solutions around a non-rotating black hole.
Solutions are for $a=0$\/ and $E=1.015$\/, with $\lam$\/ increasing
(from bottom to top in the lower diagram) as $\lam=3.0$\/, 3.25, 3.33, 3.335, 3.5, 3.8.
The sonic point is marked by $\bullet$\/.
The location of these models in the $\Es$\/-$\lams$\/-plane is marked by $\times$\/
in Fig. 1.

\vspace{1cm}
{\bf Figure 3:}
Sequence of ideal disc solutions around a rotating black hole.
Solutions are for $a=0.95$\/ and $E=1.06$\/, with $\lam$\/ increasing
(from bottom to top in the lower diagram) as $\lam=2.0$\/, 2.05, 2.08, 2.09, 2.15, 2.3.
The sonic point is marked by $\bullet$\/.
The location of these models in the $\Es$\/-$\lams$\/-plane is marked by $\times$\/
in Fig. 1.

\vspace{1cm}
{\bf Figure 4:}
Sequence of \hot ideal disc solutions around rotating black holes.
Solutions are for $E=1.2$\/ and $\lam=2.0$\/.
The Kerr parameter $a$\/ increases (from bottom to top) as
$a=0$\/, 0.2, 0.4, 0.6, 0.8, 0.9, 0.95.
The sonic point is marked by $\bullet$\/.

\vspace{1cm}
{\bf Figure 5:}
Sequence of \hot viscous disc solutions around a non-rotating black hole.
Solutions are for $a=0$\/, $\dcM=0.032$\/ and $L_0=3.5$\/, with $\alpha$\/
decreasing (from top to bottom) as $\alpha=0.045$\/, 0.04, 0.03, 0.01.
Dashed curves represent Keplerian distributions.
The sonic point is marked by $\bullet$\/.

\vspace{1cm}
{\bf Figure 6:}
Sequence of \hot viscous disc solutions around a rotating black hole.
Solutions are for $a=0.95$\/, $\dcM=0.027$\/ and $\alpha=0.022$\/,
with $L_0$\/ decreasing (from top to bottom in the $\lam$\/-diagram) as
$L_0=2.43$\/, 2.40, 2.30.
Dashed curves represent Keplerian distributions.
The sonic point is marked by $\bullet$\/.

\vspace{1cm}
{\bf Figure 7:}
Sequence of \hot viscous disc solutions around a non-rotating black hole.
Solutions are for $a=0$\/, $\dcM=0.032$\/ and $L_0=5.0$\/, with $\alpha$\/
decreasing (from top to bottom in the $\lam$\/-diagram) as $\alpha=0.008$\/, 0.006.
Dashed curves represent Keplerian distributions.
The sonic point is marked by $\bullet$\/.

\vspace{1cm}
{\bf Figure 8:}
Sequence of \hot viscous disc solutions around a non-rotating black hole.
Solutions are for $a=0$\/, $\alpha=0.001$\/ and $L_0=3.8$\/, with $\dcM$\/
decreasing (from botom to top in the $\lam$\/-diagram) as
$\dcM=0.0112$\/, 0.0028, 0.0010, 0.0004.
Dashed curves represent Keplerian distributions.
The sonic point is marked by $\bullet$\/.

\vspace{1cm}
{\bf Figure 9:}
Sequence of \cold viscous disc solutions around a non-rotating black hole.
Solutions are for $a=0$\/ and $\dcM=0.011$\/.
The upper two solutions have $\alpha=0.6$\/ and (from top to bottom in the $\lam$\/-diagram)
$L_0=1.3$\/, 1.1. 
The lower two solutions have $\alpha=0.5$\/ and
(from top to bottom in the $\lam$\/-diagram) $L_0=0.9$\/, 0.0.
Dashed curves represent Keplerian distributions.
The sonic point is marked by $\bullet$\/.

\vspace{1cm}
{\bf Figure 10:}
Sequence of \cold viscous disc solutions around a rotating black hole.
Solutions are for $a=0.95$\/, $\dcM=0.011$\/ and $L_0=1.0$\/, with
$\alpha$\/ decreasing (from top to bottom in the $\lam$\/-diagram) as
$\alpha=0.8$\/, 0.7, 0.3.
Dashed curves represent Keplerian distributions.
The sonic point is marked by $\bullet$\/.

\vspace{1cm}
{\bf Figure 11:}
Different expressions for the shear tensor component $\sigma^r_\phi$\/,
plotted for the supersonic part of a typical \cold solution 
($a=0$\/, $\alpha=0.5$\/, $\dcM=0.017$\/, $L_0=0.9$\/).
The solid line describes the full expression (\ref{eqa4}), used to calculate
the underlying disc solution. 
The short dashed line corresponds to approximation (\ref{eqa6}). 
Other curves represent expressions obtained from Eq. (\ref{eqsigmaNT73})
with $\gam$\/ and $\Omega'$\/ replaced by $\gamk$\/ and $\Omegak'$\/
(long dashed line) and with $\gamt$\/ and $\Omega'$\/ replaced by $\gamk$\/ 
and $\Omegak'$\/ (short dash-long dashed line).

\vspace{1cm}
{\bf Figure 12:}
Different expressions for the relative half-thickness $h/r$\/, 
monitored for a typical \cold solution ($a=0$\/, $\alpha=0.5$\/, $\dcM=0.017$\/, $L_0=0.9$\/).
The solid line describes the full expression (\ref{neweq6}). 
Dashed lines correspond to h/r as derived from Eq. (\ref{neweq2}), with  
$\gam$\/ instead of $\gamk$\/ (short dashed) and $\gamt$\/ instead of $\gamk$\/ (long dashed).

\vspace{1cm}
{\bf Figure 13:}
Parameter space showing contours of constant critical sound speed $\cs$\/.
The critical angular velocity $\Omegas^\pm$\/ (bottom) and
the corresponding critical specific angular momentum $\lams^\pm$\/
(top) are plotted for non-rotating ($a=0$\/; left diagrams) and for a rotating ($a=0.95$\/) black hole.
Contours correspond to (from left to right) $\cs^2=0.3$\/, 0.15, 0.08, 0.04, 0.02, 0.0.
The $\cs^2=0.0$\/ curve is identical with the Keplerian distributions $\lamk$\/, $\Omegak$\/.
In the $\Omega$\/-diagrams, short-dashed lines represent the limiting angular frequencies
$\Omega_\pm$\/ for timelike trajectories, and long-dashed lines represent the angular
velocity $\omega$\/ of the LNRF.

\vspace{1cm}
{\bf Figure 14:}
Variation of critical specific angular momenta $\lams^\pm$\/
with the location of $\rs$\/ for a set of parameters $\Es$\/.
The upper diagram is for $a=0$\/, the lower for $a=0.95$\/.
The outer dashed curves are the Keplerian distributions.
Parameters $\Es$\/ are (from the top curves to the bottom)
$\Es=0.98$\/, 1.00, 1.02, 1.03, 1.05 (dashed for $a=0$), 1.10, 1.20, 1.50 respectively.

\vspace{1cm}
{\bf Figure 15:}
Variation of the critical energy $\Es$\/ with the location
of $\rs$\/ for a set of parameters $\lams$.
The lower dashed curves are the Keplerian distributions.
For $a=0$\/ (upper diagram) parameters $\lams$\/ are
(from the top curve to the bottom) $\lam=0$, 2, 3, 3.13
($=\lam^\ast$\/; dashed), 3.4, 3.6, 3.8, 4 and 4.5 respectively.
For $a=0.95$\/ (lower diagram) parameters $\lams$\/ are
(from the top curve to the bottom) $\lams=0$, 1.9, 2, 2.03
($=\lam^\ast$\/; dashed), 2.1, 2.2, 2.3, 2.4 and 2.5 respectively.
Further plotted (dashed lines) are the curves of extrema
$\Emax$\/ and $\Emin$\/.

\end{document}